# Finer-Focused Partial Wave Spectroscopy (ff-PWS) and Detection of Cancer Stages From Human Tissue Samples.

**Dhruvil Solanki,[†] Prakash Adhikari,[†] Ishmael Apachigawo, Fatemah Alharthi, and Prabhakar Pradhan[*]**

*Department of Physics and Astronomy, Mississippi State University, Mississippi State, MS 39762, USA*

*[†]These authors contributed equally*

**Abstract:** We developed mesoscopic physics-based, further engineered finer-focused partial wave spectroscopy (ff-PWS), which can probe the precise scattering volume in cells/tissues to detect nanoscale structural alterations more effectively, even when tissue samples are embedded within different materials. Cancer progression is associated with different genetic and epigenetic events, which result in nano to microscale structural alterations in cells/tissues. However, these structural alterations in the early stage of the disease remain undetectable by conventional microscopy due to the diffraction-limited resolution. With cancer being an epidemic worldwide, techniques for accurate detection of early stages, as well as different stages of cancer, are always in demand. In this work, we first show the increase in the detection efficiency of ff-PWS relative to general PWS. Deadly cancer, such as pancreatic, prostate, breast, and colon cancer tissue microarrays (TMA) samples, paraffin-embedded, containing multiple cores of different stages for each cancer, are analyzed using this further engineered PWS technique. Using ff-PWS and commercially available TMA samples, this quantitative analysis of different cancer stages could enhance and standardize efficient, accurate cancer diagnostics and research.

## 1. Introduction

The optical detection of structural changes in biopsy samples due to progressive cancer has achieved tremendous success in detecting the stages of cancer when it develops into the late stages of a tumor **[1]**. The standard pathological method of detection of cancer stages includes microscopic examination of the morphological changes using stained biopsy samples. However, due to the diffraction-limited resolution of conventional microscopy, detecting the structural alterations in healthy tissues before the development to late stages is not very effective. These structural alterations of healthy cells/tissue are due to the rearrangements of the basic building blocks of a cell, which are macromolecules such as DNA, RNA, lipids, etc., whose size ranges from 100-200 nm, with the progression of cancer. It is recognized that genetic and epigenetic alterations occur not only at the neoplastic focus but more diffusely in the field of cancerization or field carcinogenesis. The abnormalities are present in the tissue surrounding the cancerous region or transformed cell primary tumor due to field cancerization **[2,3]**, initially at the nanoscale level before tumor formation and migration. This effect is observed in almost every type of cancer that is diagnosed in the later stages with a lower survival rate. Therefore, a susceptible optical method to detect such structural abnormalities before developing a tumor for all cancers is paramount for decreasing lethality **[4]**. Several microscopic imaging techniques were employed with time, and despite their drawbacks, significant success has been achieved **[5]**. In particular, it is now known that cancer progression is more associated with changes in mass density/refractive index fluctuations at early stages relative to changes in bulk properties of a cell and tissue. Furthermore, most used optical methods, such as phase contrast microscopy and optical coherence tomography, are still not sensitive enough to probe the nanoscale structural changes but are suitable for bulk changes. In light of this scenario, the versatile approach of using recently developed finer focus partial wave spectroscopy (PWS) to probe changes in the refractive index (RI) fluctuations using tissue microarray (TMA) samples that are embedded in paraffin could standardize the cancer research and diagnostic modalities. General PWS, developed earlier, is not very sensitive if the cells are kept within the slide and cover slip or if the tissue is embedded in other media. The newly introduced ff-PWS technique is sensitive enough to probe nanoscale refractive index fluctuation in paraffin-embedded TMA samples. The PWS combines interdisciplinary approaches of mesoscopic physics and optical imaging techniques to quantify the degree of structural disorder strength ($L_d$) based on the change in refractive index fluctuations within the cells/tissue **[6–9]**, and success has been reported already. The backscattered signal at any point within a weakly disordered medium contains spectral fluctuations proportional to the local density of macromolecules or RI fluctuations. Thus, once the RI fluctuations of the medium are known, the spatial variations of macromolecular density can be measured using the disorder strength ($L_d$) with $L_d = \langle dn^2 \rangle l_c$, where $\langle dn^2 \rangle$ is the square of *rms of* $n(r)$ and $l_c$ is the correlation length of the refractive index fluctuations **[6,7]**. This potential biomarker, the $L_d$ parameter, has shown tremendous success in distinguishing stages of cancer, drug effects in cancer treatment, and quantifying different types of abnormalities in the brain cells developed on slides **[6,8,10–12]**. In addition, structural alterations in biological cells/tissue due to cancer or any other abnormalities are quantified using mesoscopic physics-based molecular specific light localization technique in terms of the degree of disorder strength **[13–16]**.

With the availability of such a powerful approach to quantify the nanoscale structural alterations in cancerous tissue, the next challenge was using an almost identical tissue sample that allows high-throughput analysis to standardize diagnostic techniques. The recently developed commercially available paraffin-embedded TMA samples have been used, significantly facilitating and accelerating tissue analyses using this further engineered, highly sensitive ff-PWS technique. The TMA is a scientific form of paraffin-embedded condensed histopathology where the tissues are kept on a single glass slide to provide a miniature multiplex platform for the analysis of ~10 to 200 tissue samples **[17]**. Multiple cores of paraffin-embedded tissue, 5μm thickness and 1.5mm diameter, in a TMA allow high-throughput assessment of macromolecules in PWS analysis that standardize the research and diagnostic techniques, reducing the variability seen on the assay of individual samples. This uniformity of assay has reduced the drawbacks in specimen handling and their impact on data acquisition **[18]**. Besides these facts, TMA samples are easily accessible/available for scientific experiments requiring more than one similar sample and for various cancer cases. Also, we believe using this PWS technique in the TMA sample could be a new direction for exploring and studying the drug effect in cancer or any other abnormalities that are in demand. It can be noted that we want to probe changes in RI fluctuations in tissue samples that are embedded with paraffin.

In this paper, we first describe the engineering of the finer focus PWS instrument. Then, using this advanced engineered PWS, we will analyze four different paraffin-embedded deadly cancer TMA samples: *pancreatic, breast, colon,* and *prostate,* and will generalize the efficacy of developed ff-PWS to detect the cancer stages. Cancer is a common disease, and with no surprise, these four different cancers are the major cause of death in the U.S.. Because of lethality, prevalence, and almost no prominent physical change or symptoms until the later stages. We study the nanoscale structural changes in these paraffin-embedded controls and cancerous tissues by quantifying the degree of structural disorder ($L_d$) as a potential biomarker. In this sense, we explored the potential possibilities of this advanced PWS technique for TMA samples for the early to late stages of cancer detection.

## Methods

### *2.1 Experimental Setup:*

The engineered ff-PWS system used to probe the precise scattering volume developed in our laboratory is shown in Fig. 1. This is a modified, further engineered version of the original PWS with finer focusing. The details of the PWS experimental setup are presented elsewhere **[8,10,12,19]**. However, in brief, the developed, further engineered early PWS setup system consists of a low-coherence broadband stable white light source Xenon lamp (150W) reflected toward the 4f combination of lenses by a mirror and collimated. Collimated beams are reflected toward the 40X objectives and focus on the sample with the help of a highly sensitive XYZ motorized scanning stage.

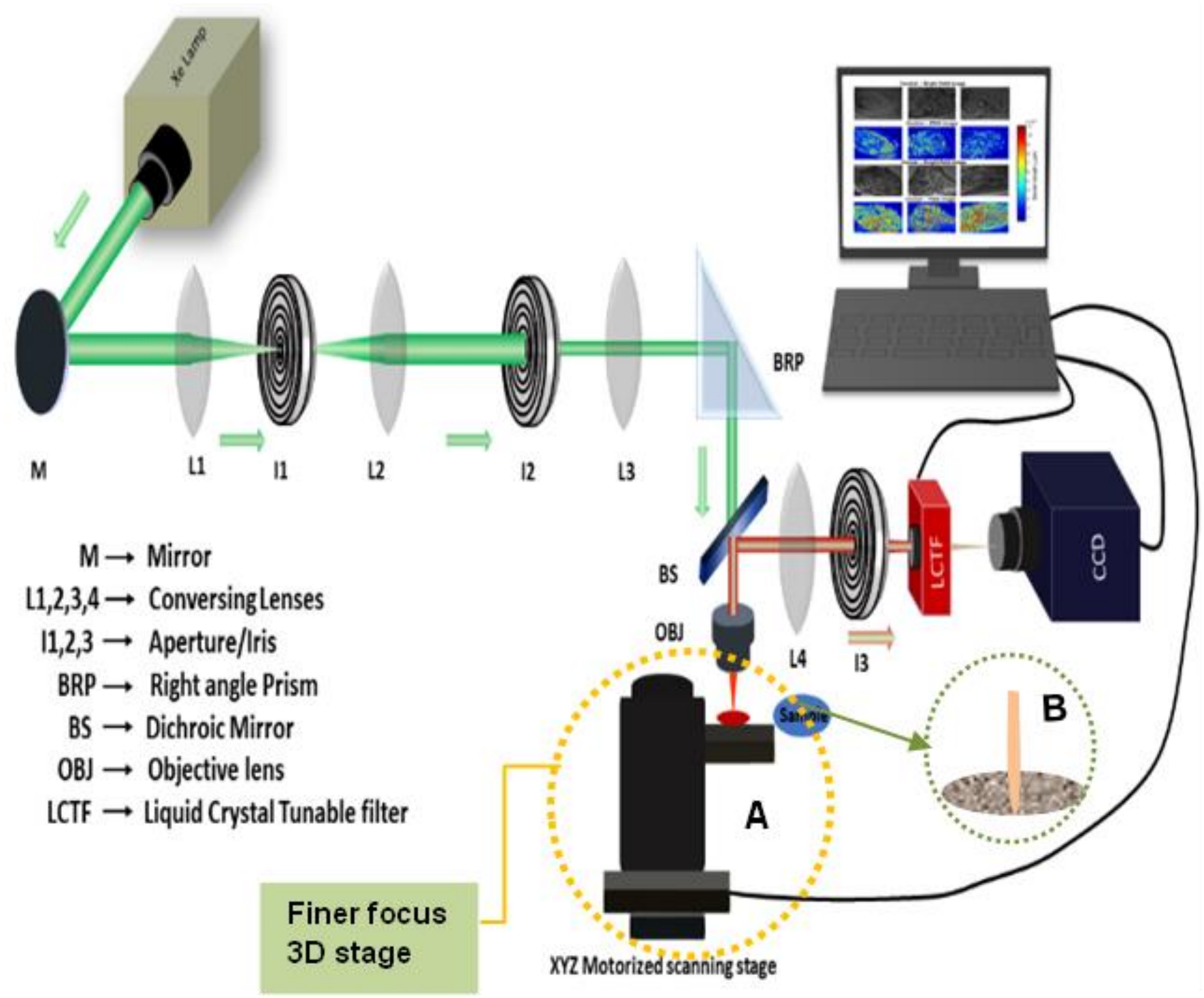

**Figure 1.** The schematic layout of the finer focusing partial wave spectroscopy (PWS) system. The collimated beam (green color) from the broadband stable white light source is focused on the sample, and the CCD collects the reflected signals (red color) from the sample for PWS analysis. Insets: (Inset A) shows the finer focus technique, and (Inset B) shows the sharp focused sample volume.

*2.2 Finer focusing scanning stage addition:*

With the earlier PWS setup, we added a finer focusing scanning stage with the following resolutions: along the XY-plane at 40 nm and the Z-axis at 100nm. The advantage of the automated finer focusing stage is that it accurately selects the scattering volume and provides more sensitivity to measuring the structural changes in the sample.

Finally, the backscattered signal from the sample is directed toward the CCD camera through the liquid crystal tunable filter (LCTF) with the help of a dichroic mirror. The LCTF has a resolution of 1nm in the visible range of light (420-730nm). Here, LCTF is coupled with a CCD camera so that the CCD records and captures images for every wavelength in the visible range. The setup's physical size is ~9 ft x 3.8 ft. CCD's image acquisition time is 100 ms. The estimated cost of the setup is $25,000. The field of view is 10 μm.

*2.3 Comparison of general and finer-focused PWS (ff-PWS) imaging systems:*

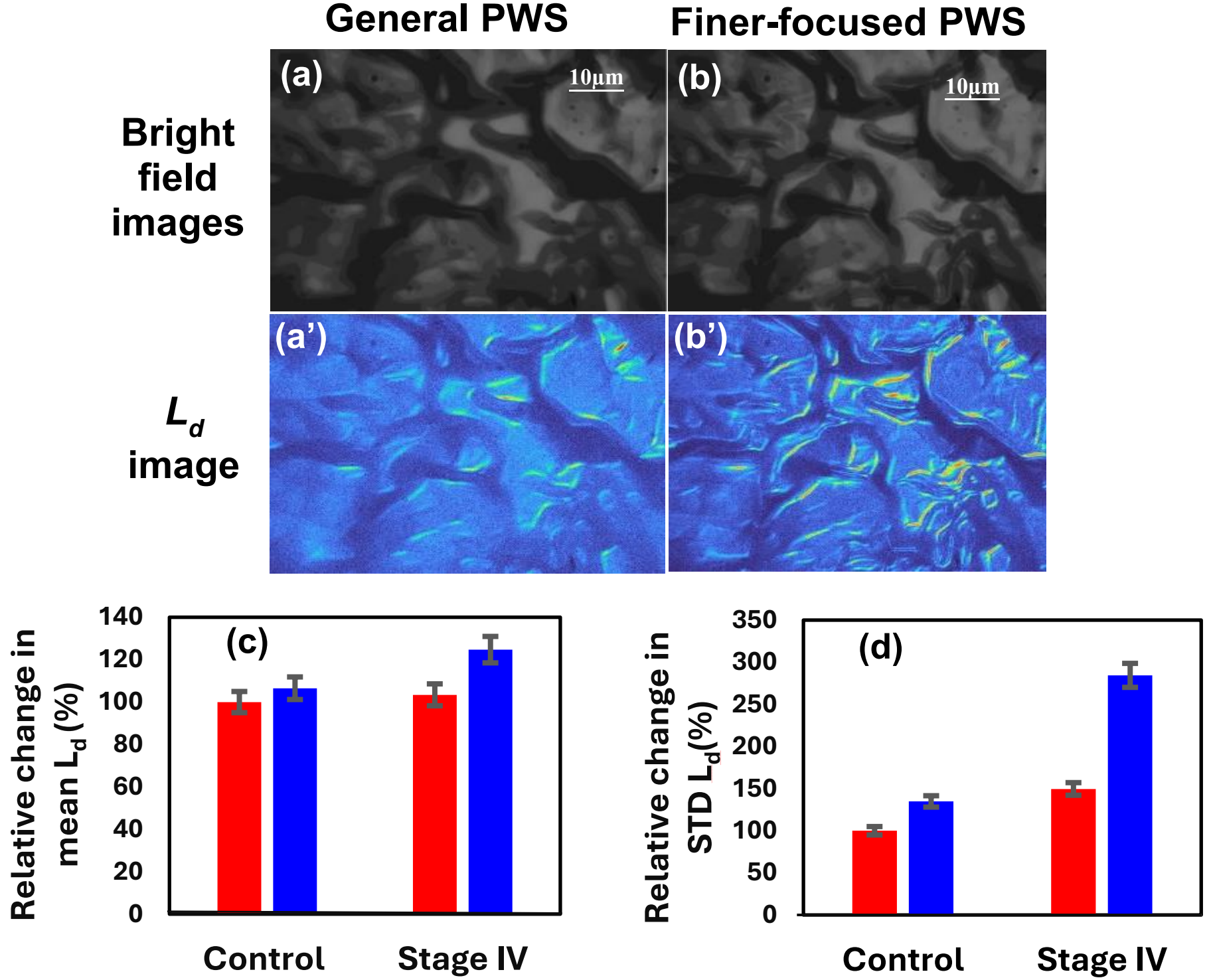

**Figure 2.** (a) and (b) are the bright field images of general PWS and ff-PWS of colon cancer TMA tissue, while (a') and (b') are their corresponding $L_d$ images. (c) and (d) is the bar graph representation of the relative change of the average and *std* of disorder strength ($L_d$) for general PWS (red) and ff-PWS (blue) with respect to general PWS control. (Students t-test *p-values* < *0.001*).

We developed a ff-PWS system to increase the sensitivity to RI fluctuations. To understand how sensitive the new system is compared to a general one, where it would not have a finer focus, an experiment was designed to compare the degree of structural disorder ($L_d$) of the system without a finer focus and with a finer focus. By comparing $L_d$, we can get insights into how sensitive ff-PWS is compared to general PWS to detect RI fluctuations, which further helps improve cancer diagnosis. In Figure 2, a, b, a', and b' are bright field images and respective $L_d$ images of colon cancer taken from general PWS and ff-PWS systems. There is an increase of 6.53% in mean $L_d$ value from general PWS (red) ff-PWS (blue) in control, whereas for stage IV colon cancer is 20.6%, as seen in (c). When we compare the standard deviation of $L_d$ values, there is an increase of 34.8% from general PWS to ff-PWS in control. The effective transverse optical resolution of the system was found by applying a mesoscopic light transport model, which came to 200 x 200 nm 1D channels within the diffraction-limited transverse size in one pixel of the sensor by a quasi-1D approximation. The transverse optical resolution in ff-PWS is 40 nm in X and Y direction and 100 nm in Z direction.

### *2.4 Calculation of the structural disorder strength ($L_d$):*

In the PWS technique, the backscattered spectrum of a weakly disordered medium is recorded in the visible range of light to quantify the degree of structural disorder strength ($L_d$) based on the refractive index fluctuations ($dn$) within the cells/tissue. It is shown that at each pixel position $(x,y)$ within the cell, the refractive index ($n$) is proportional to the local mass density. It is known that macromolecular mass density is proportional to the refractive index. Therefore, the spatial variation of macromolecular mass density at every pixel position can be expressed regarding RI fluctuations, assuming these random fluctuations are within a correlation length ($l_c$). The recorded backscattered spectrum, $R(x,y;\lambda)$, is the interference between the intercellular volume scattering with the full depth of the samples and reflection from the sample's surface. That means the measured spectra from each pixel of an image is a 1D weakly disordered medium that acts as a subset of the scattered waves.

In a 1D weakly disorder medium, the probability density distribution of reflectance R fluctuations follows a log-normal distribution for all the sample length scale $L$, which is reflected from the scattering medium. As mentioned, PWS collects the backscattered signals propagating along the 1D trajectories. Within a quasi-1D approximation, the reflection from the sample is virtually divided into many parallel channels within the diffraction-limited transverse size. Then, the degree of structural disorder strength is calculated by applying mesoscopic light transport theory and using the reflected spectral fluctuations due to the RI fluctuations. In particular, the *rms* value of the reflection intensity $\langle R(k) \rangle_{rms}$ over the visible spectrum and the spectral auto-correlation of the reflection intensity $C(\Delta k)$ for a given pixel at position $(x,y)$, are combined to define the degree of disorder strength as **[6,7,20]**:

$$L_d = \frac{B n_0^2 \langle R \rangle_{rms}}{2k^2} \frac{(\Delta k)^2}{-\ln(C(\Delta k))}\bigg|_{\Delta k \to 0} . \quad (1)$$

Where $B$ is the normalization constant, $n_0$ is the average RI of the weakly disordered medium, $k$ is the wavenumber $(k = 2\pi/\lambda)$, and $(\Delta k)^2/ln(C(\Delta k))$ is obtained by performing a linear fit of $ln(C(\Delta k))$ vs $(\Delta k)^2$ curve.

For the Gaussian color noise of the refractive index at position $r$ and $r'$, the autocorrelation of the RI fluctuations can be defined as $\langle dn(r)dn'(r')\rangle = dn^2 exp(-|r-r'|/l_c)$, then $L_d$ can be expressed as $L_d = \langle dn^2 \rangle l_c$ **[7,9]**, <> is the ensemble averaging. Therefore, the average and standard deviation (*std*) of the disorder strength quantifies the variability of the local density of intracellular material within the samples.

*2.5 Paraffin-imbedded cancer Tissue Microarrays (TMA) samples:*

TMA is a rapidly growing commercially available tissue samples delivery platform consisting of numerous cases of 5μm thick tissue cores in diameter 1.5mm placed on the same glass slide for simultaneous analysis. It generally consists of 20 to 100 cores of tissue samples. These paraffin-embedded TMA samples allow the high-throughput analysis of tissue samples since different cases of samples have exact experimental conditions and batches of reagents. TMAs are, therefore, scientific and cost-effective and offer an unprecedented degree of standardization for conducting experiments on large numbers of tissue samples on a single slide without a direct collection from the hospital for histopathological analyses and optical imaging **[21]**.

Here, using the PWS, we quantify the structural properties of paraffin-embedded TMAs from US Biomax of the pancreas (T142b), breast (BR248a), colon (T054c), and prostate (T191a) cancer samples due to progressive carcinogenesis. Each TMA consists of different cores of control and cancer stages tissue samples (5μ) thickness: control, stage I, stage II, and stage III cancer tissue samples. Each TMA's Different cores are from patients of varying ages and sexes. However, for each stage of each cancer type, different cores of similar age and sex were analyzed to calculate the degree of structural disorder.

**Results and Discussion:**

The application of the developed, highly sensitive PWS technique is to detect the nanoscale RI fluctuations. To develop a standardized diagnostic clinical research test for early and effective detection of stages of varying types of the deadliest cancers, the structural alterations were quantified using a finer focusing PWS technique of TMAs. Here, the degree of disorder strength ($L_d$) for different stages of pancreatic, breast, colon, and prostate cancer TMAs were computed and compared. The results show that structural alterations increase with the progression of cancer stages in each type of progressive cancer. Since the disorder strength ($L_d$) is the product of the variance and spatial correlation length of the RI fluctuations, $L_d = <\Delta n^2> \times l_c$, these results indicate that the progression of cancer increases more mass rearrangements/alterations as well as accumulations in the tissue results in increasing the refractive index spatial fluctuations increasing the $L_d$ value. In addition to the earlier findings, this work explores fluctuations in paraffin-embedded TMA tissue samples of four deadly cancers.

For each type of cancer, TMA samples from *pancreatic, breast, colon*, and *prostate* tissue spectroscopic images were recorded in the visible spectral range (450-700nm) of light from three different cores of the same type. From each core tissue sample, at least 5 different realizations were made for the PWS analysis. Using this powerful spectroscopic technique, tissue's pixel-wise refractive index fluctuations were computed and first represented in the 2D $L_d$ map/image. Then, the ensemble average and standard deviation (*std*) of the disorder strength ($L_d$) were computed and presented in a bar graph for all cancerous TMAs studied.

*3.1 Pancreatic cancer:*

Pancreatic cancer is a lethal condition worldwide due to poor outcomes and a rising incidence rate. It is 3rd leading cause of cancer deaths in the US, common in both men and women and often presents at an advanced stage, which contributes to a poor five-year survival rate **[22]**. Because of its physical orientation and lack of early symptoms, a better understanding of the symptoms associated with pancreatic cancer and its risk factors is essential to both health professionals and individuals. Therefore, we focus our research with added finer focusing on the existing PWS technique for early diagnosis and to understand the structural properties of pancreatic tissue with progressive carcinogenesis using TMA tissue samples. The results show that mass density or refractive index fluctuations increase with the increase in pancreatic cancer stages.

Fig. 3(a)-(b) and (a')-(b') are the representative bright-field and their corresponding $L_d$ images of normal and stage III pancreatic cancer tissue samples. The $L_d$ map, a 2D image average along the z-axis, represents the refractive index fluctuations at that point. The $L_d$ map shows that the stage III pancreatic cancer sample has more red spots, which indicates higher RI fluctuations in the color map. Further, the average and *std* of the degree of disorder strength were calculated and represented in the bar graph, Fig.

3(c)-(d). Both the average and *std* of $L_d$ value increase from normal to stage I, stage II, and stage III with a significant difference (Student's t-test *p-values* < *0.001*). Therefore, this result suggests that the developed ff-PWS technique is sensitive enough to detect nanoscale changes in RI fluctuation and distinguish the different stages of cancer in paraffin-embedded pancreatic TMA samples.

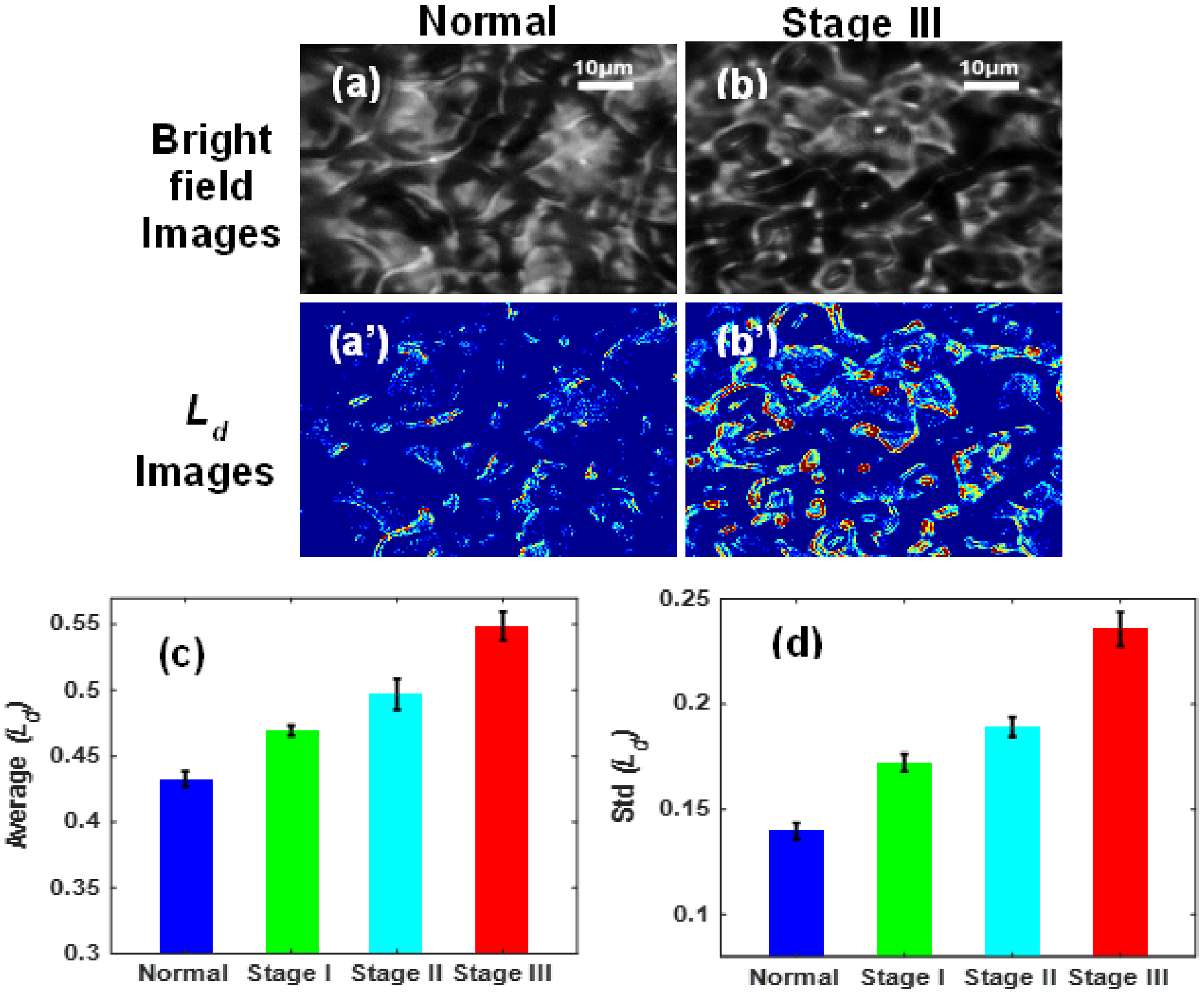

**Figure 3. Pancreatic cancer:** (a) and (b) are the representative bright field images of normal and stage III pancreatic cancer TMA tissue, while (a') and (b') are their corresponding $L_d$ images. (c) and (d) are the bar graph representing the average and *std* of disorder strength ($L_d$) for normal and different stages of pancreatic cancer TMA samples. The PWS analysis of TMA tissues shows that both the average and *std* of $L_d$ *value increase* from the normal to higher stages of pancreatic cancer. About the normal, the std of $L_d$ value of cancer stage I is 23%, stage II is 36%, and stage III is 69% higher (Students t-test *p-values* < *0.00*1, n=15).

*3.2 Breast cancer:*

Breast cancer is the most common cancer in women and a challenging cause of cancer death in the world. This metastatic cancer is transferable to different organs such as the bones, lungs, liver, and brain with almost incurability. Early diagnosis of breast cancer is only the best method of progressive carcinogenesis. The result shows that the structural disorder ($L_d$) increases significantly from the normal to stage III, as shown in Fig. 4. of prevention. Although the five-year relative survival rate of an early detected breast cancer patient is more than 80% nowadays, oncologists and scientists are still struggling to develop a technique that can identify early symptoms and distinguish the nanoscale structural changes in cells/tissues of breast cancer **[23]**. We applied finer focused PWS on breast cancer TMA to distinguish different stages of cancer. Figure 4 a, b shows bright field images of control and stage III breast cancer, whereas a', b' are their respective $L_d$ images. Significant increase in mass disorder is noticeable in c, d as cancer progresses from control.

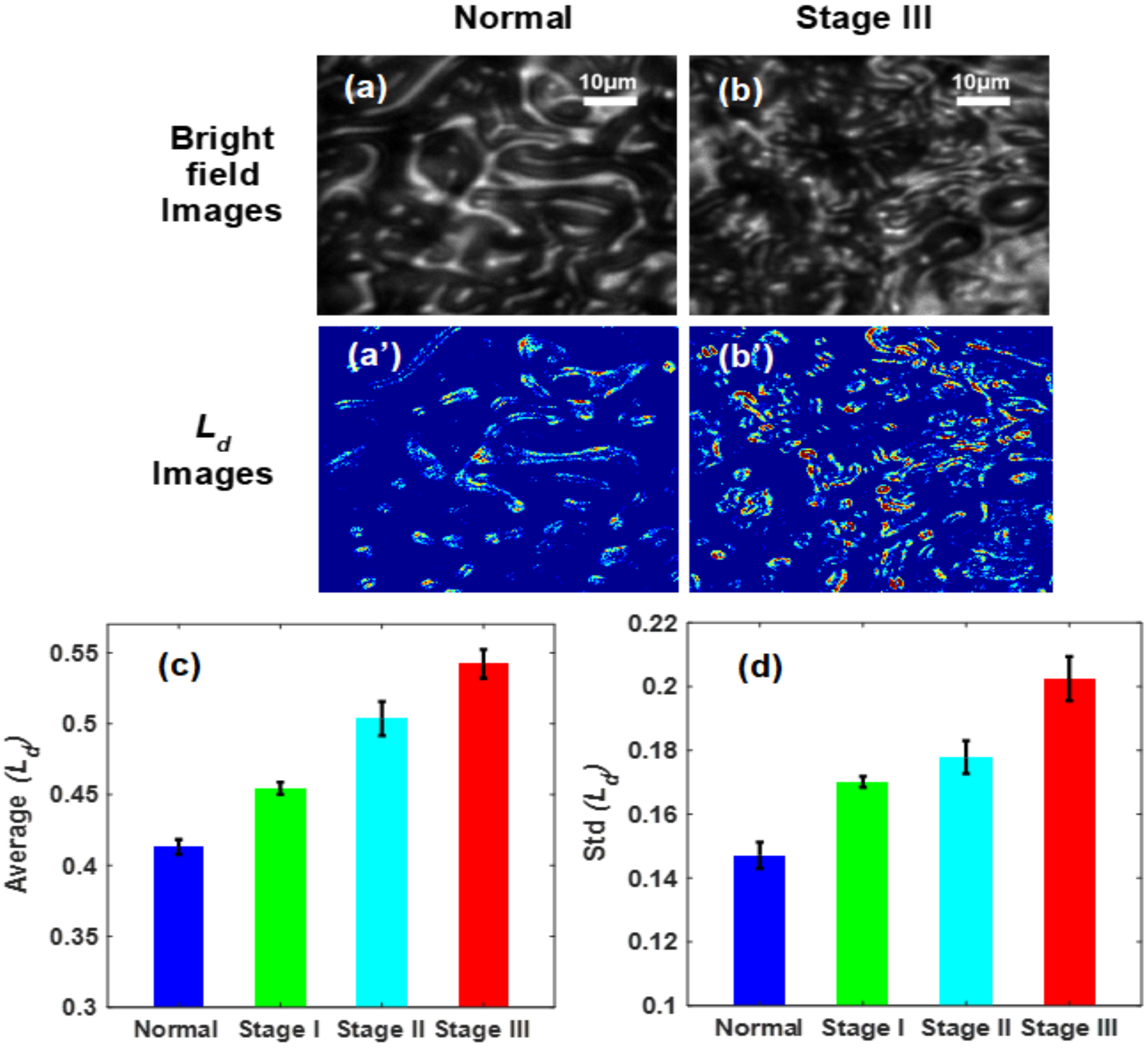

**Figure 4. Breast Cancer:** (a) and (b) are the bright field images, while (a') and (b') are the $L_d$ images of the normal and stage III breast cancer of TMA tissue, respectively. (c) and (d) are the bar graphs of the average and *std* of disorder strength ($L_d$) for the normal and different stages of breast cancer TMA. The PWS result using TMA tissue shows that the average and *std* of $L_d$ value increases from normal to higher stages of cancer tissue. About the normal, the std of $L_d$ value of cancer stage I is 16%, stage II is 21%, and stage III is 38% higher (Student t-test $p\text{-values} < 0.001$, n=15)

### *3.3 Colon cancer:*

It is considered that colon cancer is predominant among cancer mortality, accounting for nearly 10%, which is also the third leading cause of deaths related to cancer in both men and women in the US. Various factors include increasing cases of metastatic colon cancer, such as population aging, smoking, low physical activity, not having a proper diet, and obesity **[24,25]**. Laparoscopy surgery has been used extensively for the treatment of primary and metastatic colon cancer. The survival rate of patients has changed a little in modern times, where there are many advances in the medical field. So, to make a standardization in screening to detect colon cancer at early stages, we need to understand the structural changes in cells/ tissues at the local level as the cancer progresses. We found that more mass accumulated as the cancer progressed, as shown in Fig. 4 (c) and (d). There is a significant increase in mean and *std* ($L_d$) values for different stages of colon cancer compared to control, which also shows that refractive index fluctuation increases.

### *3.4 Prostate cancer:*

Prostate cancer is a common cancer among men, especially elderly ones, and 1 in 9 men will have prostate cancer during their lifespan. It is 5$^{th}$ leading cause of cancer death. Ceasing smoking, proper exercise, and weight control are good health practices that may reduce the risk of developing prostate cancer. However, remarkable progress has been achieved on characterizing risk factors and identifying therapeutic treatment options. Screening for and diagnosing the early stages of prostate cancer is still one of the most challenging issues across the globe in medicine **[23,24]**. At this point, the study of structural properties of prostate cancer tissue at the nanoscale level using a precise volume scattering technique, PWS of TMA tissue could help to characterize the structural alterations in their different stages. The results obtained using the PWS technique to study the commercially available prostate TMA are presented in Fig. 5. The representative $L_d$ images i.e. Fig. 5(a')-(b') shows that stage III Prostate cancer has more red spots in the color map because it has a higher mass density or RI fluctuations than normal. To quantify the structural abnormalities, the average and *std* of the degree of disorder strength ($L_d$) were computed and represented in the bar graph, as shown in Fig. 5(c)-(d). As can be seen from the figures, both the average and *std* of $L_d$ value increase in stage I, stage II, and stage III compared to normal. This increase in the $L_d$ is due to the increase in RI fluctuations with the increase in the stages of prostate cancer.

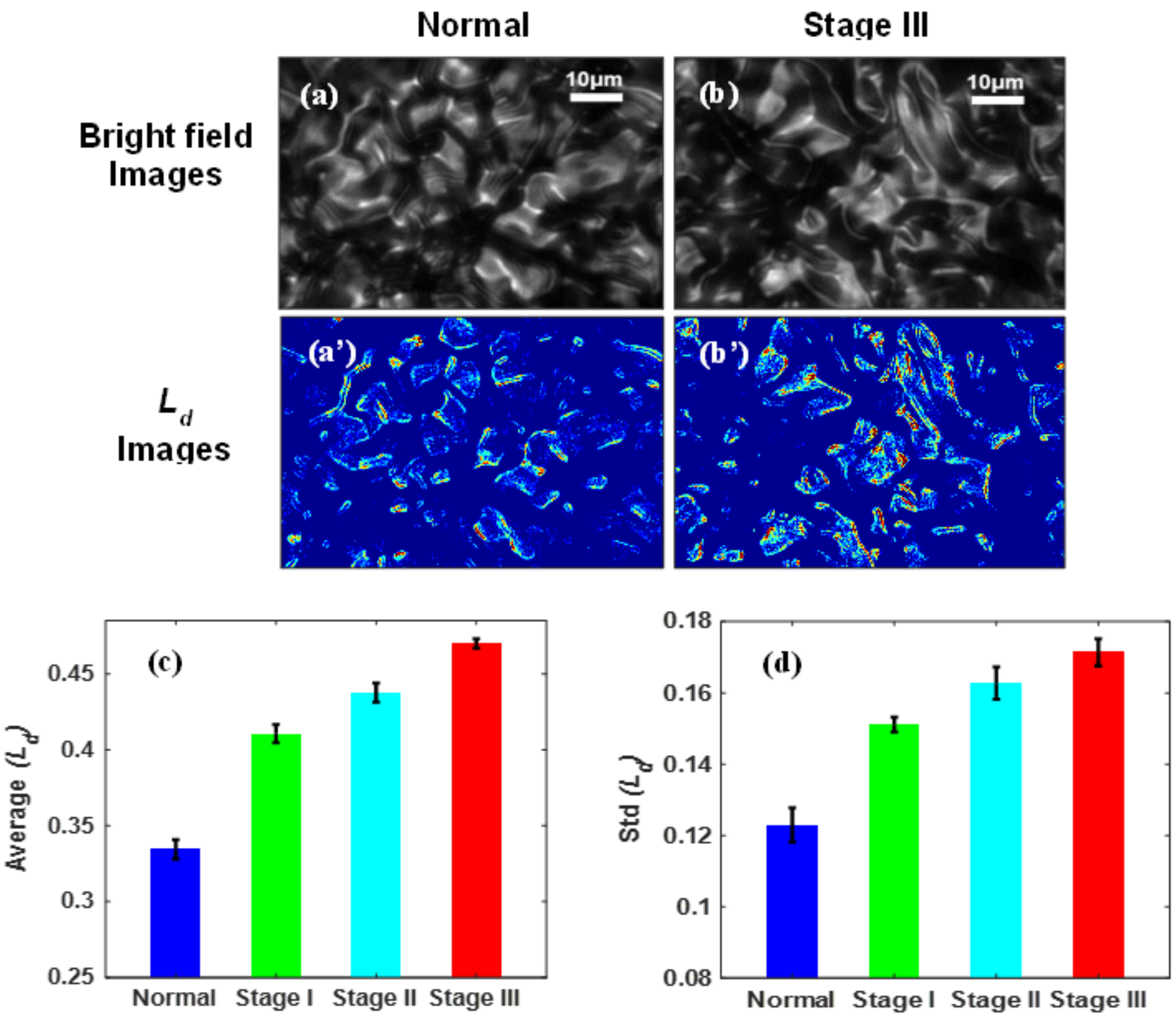

**Figure 5. Colon cancer:** (a) and (b) are the bright field images of the normal and stage III colon cancer TMA tissue and (a') and (b') are their respective $L_d$ Images. (c) and (d) bar graphs of the average and *std* of disorder strength ($L_d$) for the normal and different stages of colon cancer TMA samples. The PWS result of TMA tissue shows that the average and *std* of $L_d$ value increases from normal to higher stages of colon cancer. In reference to the normal, the std of $L_d$ value of cancer stage I is 23%, stage II is 32%, stage III is 40% higher (Student's t-test *p-values* $< 0.001$, n=15).

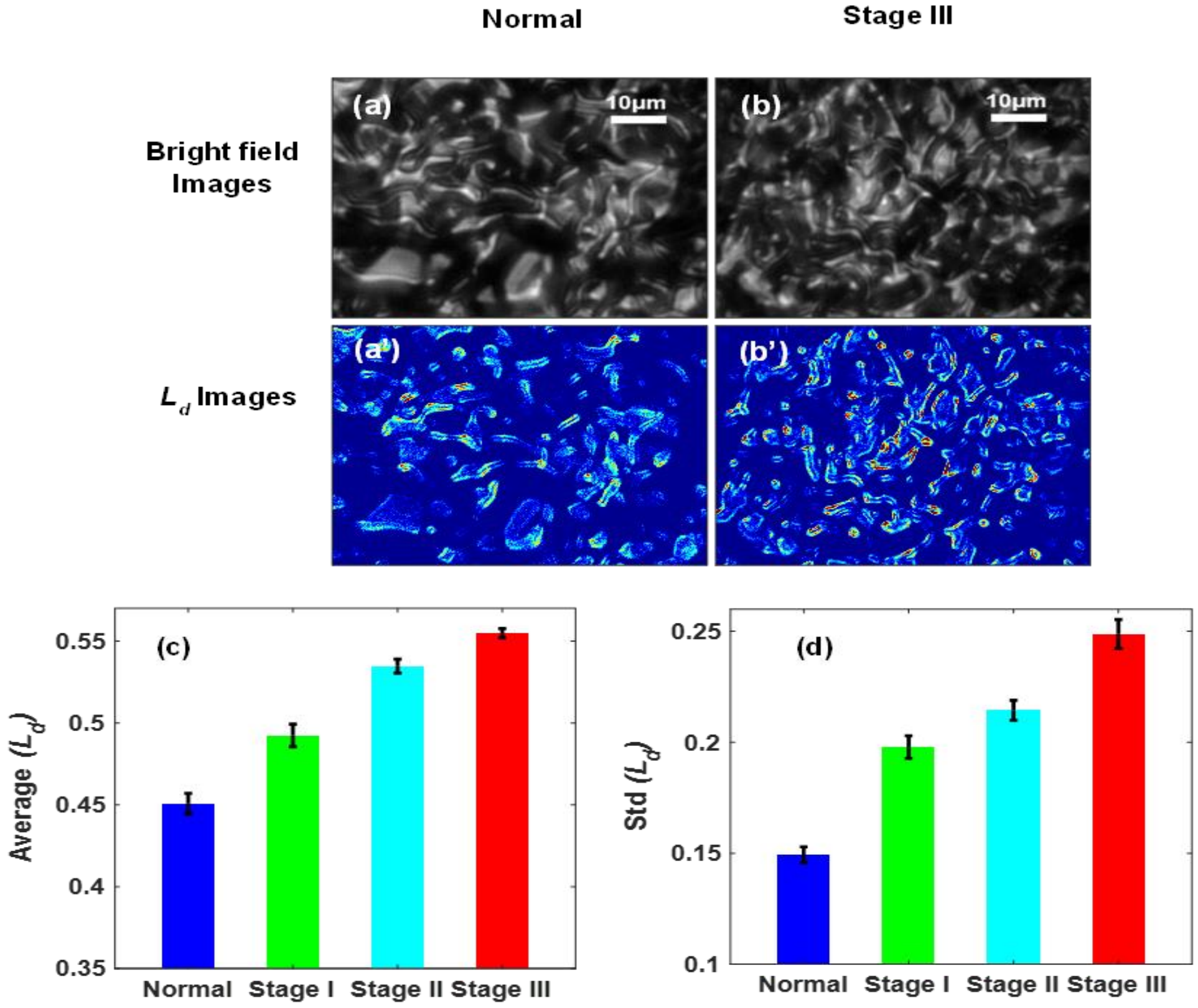

.

**Figure 6. Prostate cancer:** (a) and (b) are the representative bright field images of the normal and stage III prostate cancer TMA tissue while (a') and (b') are their corresponding $L_d$ images. (c) and (d) are bar graphs of the average and *std* of disorder strength ($L_d$) for the normal and different stages of prostate cancer TMA. The PWS analysis of TMA tissue samples show both the average and *std* $L_d$ value increases from the normal to higher stages of cancer tissue. In reference to the normal, the *std* ($L_d$ )value of cancer stage I is 32%, stage II is 46%, and stage III is 65% higher (Student's t-test *p-values* < *0.001*, n=15).

## Conclusions:

In summary, an alternative method in cancer research, applying the mesoscopic physics-based technique using commercially available TMAs, has been reported. This technique uses PWS, the recently developed finer-focusing PWS technique, which can scatter precise volume to detect the nanoscale structural changes in paraffin-embedded tissue and accurately distinguish the stages of different cancer cases. A relative study of general PWS and ff-PWS shows that ff-PWS can detect the RI fluctuations more accurately. This is because the finer focusing brings less averaging of the fluctuations. The degree of disorder strength ($L_d$) of different tissues with the clinically recorded progression of cancer stages of different TMA samples are quantified to validate the purpose method. The PWS results obtained for some of the deadliest cancer TMA samples: *pancreatic*, *breast, colon*, and *prostate* show that the average and *std* disorder strength ($L_d$) increases significantly as cancer progresses through the different stages. However, the more prominent changes were found in the degree of *std*($L_d$ ) value for all case studies. The results of the proposed method followed those of earlier studies [6,10]. Therefore, the $L_d$ parameter acts as a sensitive potential biomarker to distinguish and standardize the cancer stages, which is very important. In addition,, using the PWS technique and commercially available TMAs could provide easy and clinically accessible samples to study the drug effect in cancer treatment. In particular, this finding invites the research communities to work with a common goal to standardize early and accurate detection of the cancer stages of different deadly cancer cases using the PWS method. Further validation in distinguishing TMA tissue cancer stages can be performed using other quantitative approaches such as fractal, inverse participation ratio (IPR), etc. **[14,25,26]**. Lastly, the flexibility provided using spectroscopic techniques such as PWS in commercially available TMA samples to distinguish the early to late stages of cancer cases opens a broad way to explore and generalize the structural changes in progressive cancer for effective diagnosis and drug treatment in the future.

**Funding**. The National Institutes of Health (NIH) grant number R21CA260147 partially supported part of this work.

**Acknowledgments**. We thank Liam Elkington for his useful help with PWS experiments.

**Disclosures.** The authors declare no conflicts of interest.

**Data availability.** Data underlying the results presented in this paper are not publicly available at this time but may be obtained from the authors upon reasonable request.